\documentclass[english]{revtex4}
\usepackage[T1]{fontenc}
\usepackage[latin1]{inputenc}
\usepackage{array}
\usepackage{amsmath}
\usepackage{amssymb}

\makeatletter

\providecommand{\tabularnewline}{\\}

\usepackage{babel}
\makeatother
\begin{document}

\title{Irreducible bilinear tensorial concomitants of an arbitrary complex
bivector}

\author{T. D. Carozzi}

\email{T.Carozzi@sussex.ac.uk}

\affiliation{Space Science Centre, Sussex University, Brighton BN1 9QT, United
Kingdom}

\author{J. E. S. Bergman}

\affiliation{Swedish Institute of Space Physics, Uppsala Sweden}

\begin{abstract}
Irreducible bilinear tensorial concomitants of an arbitrary complex
antisymmetric valence-2 tensor are derived in four-dimensional spacetime.
In addition these bilinear concomitants are symmetric (or antisymmetric),
self-dual (or anti-self-dual), and hermitian forms in the antisymmetric
tensor. An important example of an antisymmetric valence-2 tensor,
or bivector, is the electromagnetic field strength tensor which ordinarily
is taken to be real-valued. In generalizing to complex-valued bivectors,
the authors find the hermitian form versions of the well-known electromagnetic
scalar invariants and stress-energy-momentum tensor, but also discover
several novel tensors of total valence 2 and 4. These tensors have
algebraic similarities to the Riemann, Weyl, and Ricci tensors.
\end{abstract}

\pacs{02.10.Xm; 03.30.+p; 03.50.De; 45.10.Na}

\keywords{Complex bivector; Irreducible tensor; Bilinear concomitant; Stress-energy-momentum
tensor}

\maketitle

\newcommand{\dual}{\vphantom{F}^{\ast}\!}

\newcommand{\blank}[1]{\phantom{#1}}

\newcommand{\CompConj}[1]{\bar{#1}}

\newcommand{\CompConjW}[1]{\overline{#1}}

\newcommand{\icu}{\mathbf{i}}

\newcommand{\selfDual}{\vphantom{F}^{{\scriptscriptstyle {+}}}\!}

\newcommand{\selfDualAnti}{\vphantom{F}^{{\scriptscriptstyle {-}}}\!}

\section{Introduction}

Bilinear, or second-order, concomitants of antisymmetric tensors of
total valence 2, also known as bivectors, are of fundamental importance
in electromagnetism and general relativity. In electromagnetism, see
\cite{Jackson99}, the electromagnetic field strength is described
by the normally real-valued tensor of total valence-2\begin{equation}
F^{\alpha\beta}=\left(\begin{array}{cccc}
0 & -E_{1} & -E_{2} & E_{3}\\
E_{1} & 0 & -B_{3} & B_{2}\\
E_{2} & B_{3} & 0 & -B_{1}\\
E_{3} & -B_{2} & B_{1} & 0\end{array}\right)\quad\alpha,\beta=0,1,2,3\end{equation}
which is antisymmetric, that is, $F^{\alpha\beta}=-F^{\beta\alpha}$.
All the bilinear tensorial concomitants of a real-valued $F^{\alpha\beta}$
of total valence two or less are well-known. They are: the two scalars\begin{align}
\mathcal{L}_{+}^{(r)}= & \frac{1}{4}F^{\mu\nu}F_{\nu\mu}\label{eq:ScalarS}\\
\mathcal{L}_{-}^{(r)}= & \frac{1}{8}F^{\mu\nu}F^{\alpha\beta}\epsilon_{\nu\mu\alpha\beta},\label{eq:ScalarP}\end{align}
where $\epsilon^{\alpha\beta\gamma\delta}$ is the Levi-Civita tensor
in four dimensions (with $\epsilon^{0123}=-1$), and the valence-2
tensor\begin{equation}
T_{(r)}^{\alpha\beta}=F^{\alpha\mu}F_{\mu}^{\blank{{\mu}}\beta}-\frac{1}{4}\mathcal{L}_{+}^{(r)}g^{\alpha\beta}\label{eq:RealEMdens}\end{equation}
where $g^{\alpha\beta}$ is the metric tensor. In electromagnetism,
(\ref{eq:ScalarS}) is the Lagrangian density of the free electromagnetic
field and (\ref{eq:RealEMdens}) is the electromagnetic stress-energy-momentum
tensor. In general relativity, the stress-energy-momentum tensor plays
a crucial role in the Einstein-Maxwell {}``electrovac'' equations,
see \cite{Penrose84}.

Usually bivectors are taken to be real-valued, and in this case the
bilinear concomitants $\mathcal{L}_{+}^{(r)}$, $\mathcal{L}_{-}^{(r)}$,
$T_{(r)}^{\alpha\beta}$ are the only possible concomitants tensors
of total valence two or less. However, if we consider the bivector
to be complex-valued, it is possible to construct an entirely new
set of bilinear tensorial concomitants which have previously not been
published. There are several reasons why one should consider arbitrary
complex bivectors. For quantum radiation, the field strength operators
of the electromagnetic field are necessarily complex and correspond
directly to electromagnetic field observables, see \cite{Glauber63};
and also for classical radiation it is convenient to treat wave-like
phenomena using complex field variables, see \cite{Wolf54}. The imaginary
part of the bivector can, for instance, be constructed by taking the
Hilbert transform of the real-valued fields, and the result is known
as the analytical signal representation. We will, however, not use
any properties particular to analytic signals, only that the bivector
is arbitrarily complex. Curiously, the term {}``bivector'' seems
to have been originally used by Gibbs \cite{Gibbs81} to denote complex-valued
3-vectors.

Thus the purpose of this paper is to explore the bilinear tensorial
concomitants of an arbitrary complex bivector. First, we construct
a set of valence-4 tensor concomitants bilinear in $F^{\alpha\beta}$
from which all others can be created through appropriate linear combinations
and contractions; these are contracted to find all tensors of total
valence 2 and 0; then we construct irreducible tensors for all valences;
and then ultimately we assemble a complete set of irreducible tensors.
All these tensor are such that they are hermitian forms in the bivector,
invariant up to a sign under a duality transformation of the bivector,
either symmetric or antisymmetric, and irreducible. Thus by construction,
these tensors have properties that match those possessed by (\ref{eq:ScalarS}),
(\ref{eq:ScalarP}), and (\ref{eq:RealEMdens}). Although we are particularly
interested in the concomitants of the Maxwell bivector, we will keep
the treatment general so it can be applied to any covariant complex
bivector. Finally examining the constructed irreducible bilinear tensor
concomitants we find that several of them are completely novel. 

This work is related to Olofsson \cite{Olofsson01} which attempts
to decompose valence-4 bilinear concomitants of complex bivectors.
An alternative approach to constructing bilinear concomitants of complex
bivectors using tensor calculus as used here, is to use spinor calculus.
A paper using a spinor approach is in preparation \cite{Sundkvist05s}.
The tensors constructed here are also indirectly related to the Riemann,
Weyl, and Ricci tensors, in that they share some algebraic properties.

\section{Criteria for the concomitants and basic assumptions\label{sec:Criteria} }

In what follows, we will consider an arbitrary (in general not self-dual)
contravariant complex bivector $F^{\alpha\beta}$. That is, we take
$F^{\alpha\beta}$ to fulfill \begin{equation}
F^{\alpha\beta}=-F^{\beta\alpha},\quad\textrm{where }F^{\alpha\beta}\in\mathbb{C}\label{eq:MainCompFdef}\end{equation}
 for all $\alpha,\beta=0,1,2,3$ (Greek indices run over 0,1,2,3 where
0 is the time-like dimension and the rest are space-like).

A bivector in spacetime can be constructed from two 3-vectors: one
being the space-time part of the corresponding valence-2 tensor, the
other the time-space part. That is, for any two complex 3-vectors\begin{align}
\mathbf{E}= & (E_{1},E_{2},E_{3})^{\mathsf{T}}\label{eq:3vecE}\\
\mathbf{B}= & (B_{1},B_{2},B_{3})^{\mathsf{T}},\label{eq:3vecB}\end{align}
where $\mathsf{T}$ denotes the transpose operator, the tensor\begin{equation}
F^{\alpha\beta}=\left(\begin{array}{cc}
0 & -\mathbf{E}^{\mathsf{T}}\\
\mathbf{E} & \mathbf{B}\times\end{array}\right)\label{eq:FinEBvec}\end{equation}
 is an arbitrary complex bivector, where $\mathbf{B}\times=\epsilon_{ijk}B_{j}$
and $\epsilon_{ijk}$ is the Levi-Civita tensor in three dimensions
(lowercase italic letters $i,j,k=1,2,3$ represent Cartesian components
in 3-space). A complex bivector is easily seen to have 12 real-valued
degrees of freedom in general.

Since $F^{\alpha\beta}$ in general has an imaginary part and we wish
to construct hermitian form concomitants, we need to consider the
complex conjugate of the bivector. We denote the complex conjugation
of a tensor $\CompConj{F}^{\alpha\beta}$, and define it as \begin{equation}
\CompConj{F}^{\alpha\beta}:=\Re\left\{ F^{\alpha\beta}\right\} -\icu\Im\left\{ F^{\alpha\beta}\right\} ,\end{equation}
where $\Re\{\cdot\}$ and $\Im\{\cdot\}$ denote the real part and
imaginary part respectively of its argument. Naturally, the complex
conjugate of the bivector is also antisymmetric so $\CompConj{F}^{\alpha\beta}=-\CompConj{F}^{\beta\alpha}$.

The (contravariant) dual of $F^{\alpha\beta}$ is defined as\begin{equation}
\dual F^{\alpha\beta}:=\frac{1}{2}\epsilon^{\alpha\beta\gamma\delta}F_{\mu\nu}=\frac{1}{2}\epsilon_{\blank{{\alpha}}\blank{{\beta}}\mu\nu}^{\alpha\beta}F^{\mu\nu}.\end{equation}
If the dual is viewed as a transform mapping bivectors to bivectors,
we can define the duality transform to be

\begin{equation}
\dual:\quad F^{\alpha\beta}\mapsto\dual F^{\alpha\beta},\;\dual F^{\alpha\beta}\mapsto-F^{\alpha\beta},\;\CompConj{F}^{\alpha\beta}\mapsto\dual\CompConj{F}^{\alpha\beta},\;\dual\CompConj{F}^{\alpha\beta}\mapsto-\CompConj{F}^{\alpha\beta},\label{eq:DualTran}\end{equation}
 which is equivalent to exchanging the bivectors 3-vector components
according to $\mathbf{E}\mapsto-\mathbf{B}$ and $\mathbf{B}\mapsto\mathbf{E}$.

As with real-valued bivectors, it is possible to construct from an
arbitrary complex bivector a bivector which is invariant up to a factor
$\pm\icu$ under the duality transform, namely\begin{align}
\selfDualAnti\mathcal{F}^{\alpha\beta}= & \left(F^{\alpha\beta}+\icu\dual F^{\alpha\beta}\right)/2\\
\selfDual\mathcal{F}^{\alpha\beta}= & \left(F^{\alpha\beta}-\icu\dual F^{\alpha\beta}\right)/2\\
\left(\CompConjW{\selfDualAnti\mathcal{F}}\right)^{\alpha\beta}= & \left(\CompConj{F}^{\alpha\beta}-\icu\dual\CompConj{F}^{\alpha\beta}\right)/2\label{eq:SelfDualnew}\\
\left(\CompConjW{\selfDual\mathcal{F}}\right)^{\alpha\beta}= & \left(\CompConj{F}^{\alpha\beta}+\icu\dual\CompConj{F}^{\alpha\beta}\right)/2,\label{eq:AntiSelfDualnew}\end{align}
see \cite{Penrose84}. The $+$ and $-$ left superscripts denote
the sign of the eigenvalue, $+\icu$ or $-\icu$ respectively, under
the duality transformation. Bivectors with eigenvalue $+\icu$ under
a duality transformation are called self-dual and those with $-\icu$
are called anti-self-dual. Compared to the case of real bivectors
there is one new self-dual bivector, Eq. (\ref{eq:SelfDualnew}),
and one new anti-self-dual bivector, Eq. (\ref{eq:AntiSelfDualnew}).
Note that contrary to the case of real $F^{\alpha\beta}$, the complex
conjugate of the self-dual bivector associated with a complex $F^{\alpha\beta}$
does not, in general, give the anti-self-dual nor vice-versa, that
is $\left(\CompConjW{\selfDual\mathcal{F}}\right)^{\alpha\beta}\neq\selfDualAnti\mathcal{F}^{\alpha\beta}$
and $\left(\CompConjW{\selfDualAnti\mathcal{F}}\right)^{\alpha\beta}\neq\selfDual\mathcal{F}^{\alpha\beta}$.

To be clear, a tensorial concomitant of $F^{\alpha\beta}$ is an algebraic
combination of $F^{\alpha\beta}$, $g^{\gamma\delta}$, or $\epsilon^{\mu\nu\alpha\beta}$,
possibly with contracted indices. A bilinear tensorial concomitant
is a tensorial concomitant that is second-order in $F^{\alpha\beta}$.
Out of the possible bilinear tensorial concomitants involving complex
bivectors we will only be interested in those that are hermitian forms
in $F^{\alpha\beta}$. By hermitian form we mean a bilinear form that
is invariant under the multiplication of $F^{\alpha\beta}$ by an
arbitrary phase factor, that is, a hermitian form in $F^{\alpha\beta}$
is unaltered under the transformations\begin{equation}
F^{\alpha\beta}\mapsto\exp(+\icu\phi)F^{\alpha\beta},\quad\CompConj{F}^{\alpha\beta}\mapsto\exp(-\icu\phi)\CompConj{F}^{\alpha\beta}.\label{eq:HermFormCond}\end{equation}
For example, tensors such as $\CompConj{F}^{\alpha\beta}F^{\gamma\delta}$
or $\CompConj{F}^{\alpha\beta}F^{\gamma\delta}\epsilon_{\gamma\delta\mu\nu}$
are hermitian form tensorial concomitants in $F^{\alpha\beta}$.

In addition, the concomitants we seek should be irreducible tensors.
By irreducible tensor we mean that the tensor cannot be decomposed
into tensors of lower total valence. For valence-2 tensors in four
dimensional space this means that the tensor should be trace-free,
that is, it should vanish when fully contracted with the metric tensor
\begin{equation}
T_{\mu}^{\blank{\mu}\mu}=0.\label{eq:2Traceless}\end{equation}
For valence-4 tensors in four dimensions, irreducibility means that
the tensor should be trace-free over all pairs of indices and in addition
vanish when fully contracted with the Levi-Civita tensor. Thus, for
an arbitrary valence-4 tensor $\mathcal{T}^{\alpha\beta\gamma\delta}$
to be irreducible it must satisfy all of the following conditions

\begin{align}
T_{\mu}^{\blank{\mu}\mu\alpha\beta}=T_{\mu}^{\blank{\mu}\alpha\mu\beta}=T_{\blank{\alpha}\mu}^{\alpha\blank{\mu}\mu\beta}=T_{\blank{\alpha}\mu}^{\alpha\blank{\mu}\beta\mu}=T_{\blank{\alpha}\blank{\beta}\mu}^{\alpha\beta\blank{\mu}\mu}=T_{\mu}^{\blank{\mu}\alpha\beta\mu} & =0\label{eq:4Traceless}\\
\epsilon_{\mu\nu\alpha\beta}T^{\mu\nu\alpha\beta} & =0\label{eq:4OrthPseu}\\
\epsilon_{\alpha\mu\nu\gamma}T^{\beta\mu\nu\gamma}=\epsilon_{\alpha\mu\nu\gamma}T^{\mu\beta\nu\gamma}=\epsilon_{\alpha\mu\nu\gamma}T^{\mu\nu\beta\gamma}=\epsilon_{\alpha\mu\nu\gamma}T^{\mu\nu\gamma\beta} & =0.\label{eq:4Irred13}\end{align}
To be precise, the tensors we seek are actually real-irreducible tensors
or Cartesian tensors as they are sometimes known, see \cite{Fano59}.
By this we mean that the tensors fulfill the irreducibility conditions
above and that their basis vectors are real-valued. In what follows,
we will sometimes simply use the term {}``irreducible tensor'' since
we will not be dealing with complex base vectors, or spherical tensors
as they are also known.

In addition to the above mentioned criteria, the sought after tensorial
concomitants should also be either symmetric or antisymmetric.

Let us recapitulate the objective of this paper: we seek bilinear
concomitants of an arbitrary complex bivector $F^{\alpha\beta}$ that
are

\begin{itemize}
\item hermitian forms in the components of $F^{\alpha\beta}$: invariant
under (\ref{eq:HermFormCond}),
\item (anti)-self-dual: invariant up to sign under (\ref{eq:DualTran}),
\item symmetric or anti-symmetric,
\item irreducible: fulfilling either (\ref{eq:2Traceless}) or (\ref{eq:4Traceless}),
(\ref{eq:4OrthPseu}), (\ref{eq:4Irred13}).
\end{itemize}
These conditions have been chosen so that the constructed bilinear
concomitants of complex $F^{\alpha\beta}$ are consistent with the
properties possessed by the well-known bilinear concomitants of real
$F^{\alpha\beta}$.

Our plan on how to construct the concomitants is as follows: we take
all possible hermitian form outer products of the four self-dual tensors,
$\selfDualAnti\mathcal{F}_{\gamma\delta}$, $\selfDual\mathcal{F}_{\gamma\delta}$,
$\CompConjW{\left(\selfDualAnti\mathcal{F}\right)}_{\alpha\beta}$
and $\CompConjW{\left(\selfDual\mathcal{F}\right)}_{\alpha\beta}$,
which constitute all possible valence-4 tensor combinations; from
these tensors we take all possible contractions over two indices to
obtain all valence-2 concomitants; and then we contracted over the
finally pair of indices to obtain all the scalar concomitants. From
these tensors of total valence 0, 2, and 4 we construct irreducible
tensors which are finally assembled into an irreducible tensorial
set.

As for the metric, we assume throughout that the metric tensor is
real-valued and symmetric and that in a local coordinate system the
metric can be set to the Lorentzian metric for which $g^{\alpha\beta}=\mathrm{diag}(+1,-1,-1,-1)$.

\section{Hermitian form self dual concomitants of valence four}

We seek to construct bilinear concomitants of the complex bivector
$F_{\alpha\beta}$ which are hermitian forms and are invariant up
to a sign under the duality transformation of the bivector. To this
end we take all bilinear combinations of the complex conjugate self-dual
bivectors $\CompConjW{\left(\selfDualAnti\mathcal{F}\right)}_{\alpha\beta}$
and $\CompConjW{\left(\selfDual\mathcal{F}\right)}_{\alpha\beta}$
with the self-dual bivectors $\selfDualAnti\mathcal{F}_{\gamma\delta}$
and $\selfDual\mathcal{F}_{\gamma\delta}$. This gives the four combinations

\begin{align}
4\CompConjW{\left(\selfDualAnti\mathcal{F}\right)}_{\alpha\beta}\selfDualAnti\mathcal{F}_{\gamma\delta}= & \CompConj{F}_{\alpha\beta}F_{\gamma\delta}+\dual\CompConj{F}_{\alpha\beta}\dual F_{\gamma\delta}+\icu\left(\CompConj{F}_{\alpha\beta}\dual F_{\gamma\delta}-\dual\CompConj{F}_{\alpha\beta}F_{\gamma\delta}\right)\label{eq:SpinorMM}\\
4\CompConjW{\left(\selfDual\mathcal{F}\right)}_{\alpha\beta}\selfDual\mathcal{F}_{\gamma\delta}= & \CompConj{F}_{\alpha\beta}F_{\gamma\delta}+\dual\CompConj{F}_{\alpha\beta}\dual F_{\gamma\delta}-\icu\left(\CompConj{F}_{\alpha\beta}\dual F_{\gamma\delta}-\dual\CompConj{F}_{\alpha\beta}F_{\gamma\delta}\right)\label{eq:SpinorPP}\\
4\CompConjW{\left(\selfDualAnti\mathcal{F}\right)}_{\alpha\beta}\selfDual\mathcal{F}_{\gamma\delta}= & \CompConj{F}_{\alpha\beta}F_{\gamma\delta}-\dual\CompConj{F}_{\alpha\beta}\dual F_{\gamma\delta}-\icu\left(\CompConj{F}_{\alpha\beta}\dual F_{\gamma\delta}+\dual\CompConj{F}_{\alpha\beta}F_{\gamma\delta}\right)\label{eq:SpinorMP}\\
4\CompConjW{\left(\selfDual\mathcal{F}\right)}_{\alpha\beta}\selfDualAnti\mathcal{F}_{\gamma\delta}= & \CompConj{F}_{\alpha\beta}F_{\gamma\delta}-\dual\CompConj{F}_{\alpha\beta}\dual F_{\gamma\delta}+\icu\left(\CompConj{F}_{\alpha\beta}\dual F_{\gamma\delta}+\dual\CompConj{F}_{\alpha\beta}F_{\alpha\delta}\right).\label{eq:SpinorPM}\end{align}
These are all tensors of valence-4, self-dual, and manifestly hermitian
forms.

However, in order to simplify the proceeding and ultimately obtain
the complex analogs of the well known bilinear concomitants of real-valued
bivectors, we instead take quantities proportional to the sum and
difference of (\ref{eq:SpinorMM}) and (\ref{eq:SpinorPP}), and proportional
to the sum and difference of (\ref{eq:SpinorMP}) and (\ref{eq:SpinorPM}).
Specifically, we introduce the following four tensors 

\begin{align}
T'_{\alpha\beta\gamma\delta}:= & \CompConjW{\left(\selfDualAnti\mathcal{F}\right)}_{\alpha\beta}\selfDualAnti\mathcal{F}_{\gamma\delta}+\CompConjW{\left(\selfDual\mathcal{F}\right)}_{\alpha\beta}\selfDual\mathcal{F}_{\gamma\delta}=\left(\CompConj{F}_{\alpha\beta}F_{\gamma\delta}+\dual\CompConj{F}_{\alpha\beta}\dual F_{\gamma\delta}\right)/2\\
Q'_{\alpha\beta\gamma\delta}:= & \CompConjW{\left(\selfDualAnti\mathcal{F}\right)}_{\alpha\beta}\selfDualAnti\mathcal{F}_{\gamma\delta}-\CompConjW{\left(\selfDual\mathcal{F}\right)}_{\alpha\beta}\selfDual\mathcal{F}_{\gamma\delta}=\icu\left(\CompConj{F}_{\alpha\beta}\dual F_{\gamma\delta}-\dual\CompConj{F}_{\alpha\beta}F_{\gamma\delta}\right)/2\\
D'_{\alpha\beta\gamma\delta}:= & \CompConjW{\left(\selfDualAnti\mathcal{F}\right)}_{\alpha\beta}\selfDual\mathcal{F}_{\gamma\delta}+\CompConjW{\left(\selfDual\mathcal{F}\right)}_{\alpha\beta}\selfDualAnti\mathcal{F}_{\gamma\delta}=\left(\CompConj{F}_{\alpha\beta}F_{\gamma\delta}-\dual\CompConj{F}_{\alpha\beta}\dual F_{\gamma\delta}\right)/2\\
X'_{\alpha\beta\gamma\delta}:= & \icu\left(\CompConjW{\left(\selfDualAnti\mathcal{F}\right)}_{\alpha\beta}\selfDual\mathcal{F}_{\gamma\delta}-\CompConjW{\left(\selfDual\mathcal{F}\right)}_{\alpha\beta}\selfDualAnti\mathcal{F}_{\gamma\delta}\right)=\left(\CompConj{F}_{\alpha\beta}\dual F_{\gamma\delta}+\dual\CompConj{F}_{\alpha\beta}F_{\gamma\delta}\right)/2.\end{align}

These valence-4 tensors can be expressed in terms of the two 3-vector
components of the bivector, $\mathbf{E}$ and $\mathbf{B}$, by using
so-called bivector indexing, see \cite{Hall02}. A bivector index
is denoted with uppercase roman letters $A$ and $B$ which run through
values 1 to 6. $F^{A}$ is used to represent the tensor component
$F^{\alpha\beta}$ where the index mapping $A\leftrightarrow[\alpha\beta]$
is taken to be $1\leftrightarrow[10]$, $2\leftrightarrow[20]$ ,
$3\leftrightarrow[30]$, $4\leftrightarrow[32]$, $5\leftrightarrow[13]$,
$6\leftrightarrow[21]$. This maps the valence-2 antisymmetric tensor,
as given in (\ref{eq:FinEBvec}), according to \begin{equation}
F^{\alpha\beta}=\left(\begin{array}{cc}
0 & -\mathbf{E}^{\mathsf{T}}\\
\mathbf{E} & \mathbf{B}\times\end{array}\right)\leftrightarrow\left(\begin{array}{c}
\mathbf{E}\\
\mathbf{\mathbf{B}}\end{array}\right)=F^{A}\end{equation}
where the vector with six components on the left-hand side, $F^{A}$,
is known as a sixtor, see \cite{Sexl01}. The valence-4 tensors can
therefore be written in terms of the two 3-vectors (\ref{eq:3vecE})
and (\ref{eq:3vecB}) using the matrices\begin{align}
T'^{\alpha\beta\gamma\delta}\leftrightarrow & T'^{AB}=\frac{1}{2}\left(\begin{array}{cc}
\CompConj{\mathbf{E}}\otimes\mathbf{E}+\mathbf{\CompConj{B}}\otimes\mathbf{B} & \mathbf{\CompConj{E}}\otimes\mathbf{B}-\mathbf{\CompConj{B}}\otimes\mathbf{E}\\
-\left(\CompConj{\mathbf{E}}\otimes\mathbf{B}-\mathbf{\CompConj{B}}\otimes\mathbf{E}\right) & \mathbf{\CompConj{E}}\otimes\mathbf{E}+\mathbf{\CompConj{B}}\otimes\mathbf{B}\end{array}\right)\end{align}
\begin{align}
Q'^{\alpha\beta\gamma\delta}\leftrightarrow & Q'^{AB}=\frac{\icu}{2}\left(\begin{array}{cc}
-\left(\mathbf{\CompConj{E}}\otimes\mathbf{B}-\mathbf{\CompConj{B}}\otimes\mathbf{E}\right) & \CompConj{\mathbf{E}}\otimes\mathbf{E}+\mathbf{\CompConj{B}}\otimes\mathbf{B}\\
-\left(\mathbf{\CompConj{E}}\otimes\mathbf{E}+\mathbf{\CompConj{B}}\otimes\mathbf{B}\right) & -\left(\mathbf{\CompConj{E}}\otimes\mathbf{B}-\CompConj{\mathbf{B}}\otimes\mathbf{E}\right)\end{array}\right)\end{align}
\begin{align}
D'^{\alpha\beta\gamma\delta}\leftrightarrow & D'^{AB}=\frac{1}{2}\left(\begin{array}{cc}
\mathbf{\CompConj{E}}\otimes\mathbf{E}-\mathbf{\CompConj{B}}\otimes\mathbf{B} & \mathbf{\CompConj{E}}\otimes\mathbf{B}+\mathbf{\CompConj{B}}\otimes\mathbf{E}\\
\CompConj{\mathbf{E}}\otimes\mathbf{B}+\mathbf{\CompConj{B}}\otimes\mathbf{E} & -\left(\CompConj{\mathbf{E}}\otimes\mathbf{E}-\CompConj{\mathbf{B}}\otimes\mathbf{B}\right)\end{array}\right)\end{align}
\begin{align}
X'^{\alpha\beta\gamma\delta}\leftrightarrow & X'^{AB}=\frac{1}{2}\left(\begin{array}{cc}
-\left(\mathbf{\CompConj{E}}\otimes\mathbf{B}+\mathbf{\CompConj{B}}\otimes\mathbf{E}\right) & \mathbf{\CompConj{E}}\otimes\mathbf{E}-\mathbf{\CompConj{B}}\otimes\mathbf{B}\\
\CompConj{\mathbf{E}}\otimes\mathbf{E}-\mathbf{B}\otimes\mathbf{B} & \mathbf{\CompConj{E}}\otimes\mathbf{B}+\mathbf{\CompConj{B}}\otimes\mathbf{E}\end{array}\right)\end{align}
where $\CompConj{\mathbf{E}}\otimes\mathbf{E}=\CompConj{E}_{i}E_{j}$
is the outer or tensor product between $\CompConj{\mathbf{E}}$ and
$\mathbf{E}$. As is clear from these sixtor matrices, the four valence-4
tensors $T'^{\alpha\beta\gamma\delta}$, $Q'^{\alpha\beta\gamma\delta}$,
$D'^{\alpha\beta\gamma\delta},$ and $X'^{\alpha\beta\gamma\delta}$
together represent all hermitian forms of valence-4 in the components
of the bivector. Each of these tensors has 18 independent components.
In fact, the tensor pair $T'^{\alpha\beta\gamma\delta}$ and $Q'^{\alpha\beta\gamma\delta}$
or the pair $D'^{\alpha\beta\gamma\delta}$ and $X'^{\alpha\beta\gamma\delta}$
on their own constitute an independent and complete set of 36 hermitian
forms in the components of the complex bivector counting real and
imaginary parts separately. This corresponds to the 36 possible hermitian
forms in the components of the complex bivector counting real and
imaginary parts separately. 

The tensors defined above, $T'^{\alpha\beta\gamma\delta}$, $Q'^{\alpha\beta\gamma\delta}$,
$D'^{\alpha\beta\gamma\delta}$, $X'^{\alpha\beta\gamma\delta}$,
all have the following properties

\begin{align}
 & T'^{\alpha\beta\gamma\delta}=-T'^{\beta\alpha\gamma\delta}=-T'^{\alpha\beta\delta\gamma}=T'^{\beta\alpha\delta\gamma}\label{eq:4symRelAntiSym}\\
 & T'^{\alpha\beta\gamma\delta}=\CompConj{T}'^{\gamma\delta\alpha\beta}\label{eq:4symRelHerm}\end{align}
where we have used $T'^{\alpha\beta\gamma\delta}$ as an example for
any of the four tensors.

\section{Valence two concomitants of self dual bivectors}

Having derived the general valence-4 tensors, we can now construct
valence-2 tensors by simply contracting the valence-4 tensors over
pairs of indices. Six such contractions are possible, but from the
symmetry relations (\ref{eq:4symRelAntiSym}) it is clear that the
contraction between the indices (1,4), (2,3), (1,3), and (2,4) are
all the same up to a sign. Furthermore, the contraction between indices
(1,2) and (3,4) are zero. Therefore the all non-vanishing contractions
are, up to a sign, given by 

\begin{align}
T^{\alpha\beta}:= & T'^{\alpha\mu\nu\beta}g_{\mu\nu}=\left(\CompConj{F}_{\blank{\alpha}\mu}^{\alpha}F^{\mu\beta}+\dual\CompConj{F}_{\blank{\alpha}\mu}^{\alpha}\dual F^{\mu\beta}\right)/2\\
Q^{\alpha\beta}:= & Q'^{\alpha\mu\nu\beta}g_{\mu\nu}=\icu\left(\CompConj{F}_{\blank{\alpha}\mu}^{\alpha}\dual F^{\mu\beta}-\dual\CompConj{F}_{\blank{\alpha}\mu}^{\alpha}F^{\mu\beta}\right)/2\\
D'^{\alpha\beta}:= & D'^{\alpha\mu\nu\beta}g_{\mu\nu}=\left(\CompConj{F}_{\blank{\alpha}\mu}^{\alpha}F^{\mu\beta}-\dual\CompConj{F}_{\blank{\alpha}\mu}^{\alpha}\dual F^{\mu\beta}\right)/2\\
X'^{\alpha\beta}:= & X'^{\alpha\mu\nu\beta}g_{\mu\nu}=\left(\CompConj{F}_{\blank{\alpha}\mu}^{\alpha}\dual F^{\mu\beta}+\dual\CompConj{F}_{\blank{\alpha}\mu}^{\alpha}F^{\mu\beta}\right)/2.\end{align}
All these tensors have total valence two. Furthermore, $T^{\alpha\beta}$
and $Q^{\alpha\beta}$ are symmetric,\begin{align}
T^{\alpha\beta}= & T^{\beta\alpha}\\
Q^{\alpha\beta}= & Q^{\beta\alpha},\end{align}
real-valued\begin{align}
\Im\left\{ T^{\alpha\beta}\right\} = & 0\\
\Im\left\{ Q^{\alpha\beta}\right\} = & 0\end{align}
and trace-free as will be shown in the next section. $D'^{\alpha\beta}$
and $X'^{\alpha\beta}$, on the other hand, are neither symmetric
nor antisymmetric, have both real and imaginary parts, and have are
not trace-free. Hence $T^{\alpha\beta}$ and $Q^{\alpha\beta}$ irreducible
while $D'^{\alpha\beta}$ and $X'^{\alpha\beta}$ are not. This is
the reason for the prime accents ($'$) on $D'^{\alpha\beta}$ and
$X'^{\alpha\beta}$: they mark the fact that these tensors are not
irreducible.

In a local coordinate system, we can set the metric tensor to be Lorentzian.
In this case the two valence-2 tensors $T^{\alpha\beta}$ and $Q^{\alpha\beta}$
can be written in terms of the two complex 3-vector components of
the bivector as follows \begin{align}
T^{00}= & \left(|\mathbf{E}|^{2}+|\mathbf{B}|^{2}\right)/2\\
T^{i0}= & \Re\left\{ \CompConj{\mathbf{E}}\times\mathbf{B}\right\} \\
T^{ij}= & -\Re\left\{ \CompConj{\mathbf{E}}\otimes\mathbf{E}+\CompConj{\mathbf{B}}\otimes\mathbf{B}\right\} +T^{00}\mathbf{1}_{3}\\
T^{ji}= & T^{ij}\end{align}
\begin{align}
Q^{00}= & \Im\left\{ \CompConj{\mathbf{E}}\cdot\mathbf{B}\right\} \\
Q^{i0}= & \frac{\icu}{2}\left(\CompConj{\mathbf{E}}\times\mathbf{E}+\CompConj{\mathbf{B}}\times\mathbf{B}\right)\\
Q^{ij}= & -\Im\left\{ \CompConj{\mathbf{E}}\otimes\mathbf{B}-\CompConj{\mathbf{B}}\otimes\mathbf{E}\right\} +Q^{00}\mathbf{1}_{3}\\
Q^{ji}= & Q^{ij}\end{align}
where $\mathbf{1}_{3}$ is the unity valence-2 tensor in 3-space. 

Physically, if we take $F^{\alpha\beta}$ to be the electromagnetic
field strength tensor, then the symmetric tensor $T^{\alpha\beta}/4\pi$
can be interpreted as the hermitian form generalization of the stress-energy-momentum
tensor in Gaussian units, see \cite{Jackson99}. Parts of this hermitian
generalization can be found in the time-harmonic analysis of electromagnetic
energy such as in the complex Poynting theorem, see \cite{Jackson99}.
The other valence-2 tensors lack, as yet, a specific physical interpretation.
Having said that, the real-valued 3-vector in the time-space components
of $Q^{\alpha\beta}$ , namely $Q^{i0}=\frac{\icu}{2}\left(\CompConj{\mathbf{E}}\times\mathbf{E}+\CompConj{\mathbf{B}}\times\mathbf{B}\right)$,
is proportional to the vector mentioned in the beginning of Lipkin
\cite{Lipkin64} and related to the vector defined in equation (17)
in Carozzi \cite{Carozzi00}. Based on this relationship one can argue
that $Q^{\alpha\beta}$ is the spin-weighted stress-energy-momentum
tensor of the electromagnetic field.

\section{Scalar concomitants}

The valence-0, or scalar, concomitants can now be found by contracting
the valence-2 tensors. Of the two possible ways of contracting both
are the same since we are assuming the metric tensor is symmetric.
Thus all possible traces are given by

\begin{align}
0= & T^{\mu\nu}g_{\mu\nu}=\left(\CompConj{F}_{\mu\nu}F^{\nu\mu}+\dual\CompConj{F}_{\mu\nu}\dual F^{\nu\mu}\right)/2\\
0= & Q^{\mu\nu}g_{\mu\nu}=\icu\left(\CompConj{F}_{\mu\nu}\dual F^{\nu\mu}-\dual\CompConj{F}_{\mu\nu}F^{\nu\mu}\right)/2\\
\mathcal{L}_{+}= & D'^{\mu\nu}g_{\mu\nu}/4=\left(\CompConj{F}_{\mu\nu}F^{\nu\mu}-\dual\CompConj{F}_{\mu\nu}\dual F^{\nu\mu}\right)/2\\
\mathcal{L}_{-}= & X'^{\mu\nu}g_{\mu\nu}/4=\left(\CompConj{F}_{\mu\nu}\dual F^{\nu\mu}+\dual\CompConj{F}_{\mu\nu}F^{\nu\mu}\right)/2.\end{align}
Thus, the tensors $T^{\alpha\beta}$ and $Q^{\alpha\beta}$ are trace-free
while the trace of $D'^{\alpha\beta}$ and $X'^{\alpha\beta}$ leads
to two scalars which are the hermitian form generalization of the
well-known scalar invariants, (\ref{eq:ScalarS}) and (\ref{eq:ScalarP}).

In a local coordinate system the two scalars $\mathcal{L}_{+}$ and
$\mathcal{L}_{-}$ can be written in terms of the 3-vector components
of the bivector as follows

\begin{align}
\mathcal{L}_{+}= & \left(|\mathbf{E}|^{2}-|\mathbf{B}|^{2}\right)/2\label{eq:Ls}\\
\mathcal{L}_{-}= & -\Re\left\{ \mathbf{\CompConj{E}}\cdot\mathbf{B}\right\} .\label{eq:Lp}\end{align}

When the bivector represents a complex-valued electromagnetic field,
$\mathcal{L}_{+}/4\pi$ is the hermitian form generalization of the
flat vacuum electromagnetic field Lagrangian density in Gaussian units
and $\mathcal{L}_{-}$ is the hermitian form version of the pseudo-scalar
invariant of the electromagnetic field in flat spacetime. Their subscript
sign indicates their eigenvalue under the duality transformation which
is either $+1$ or $-1$.

\section{Irreducible concomitants of valence two}

As mentioned in the previous section, the tensors $T^{\alpha\beta}$
and $Q^{\alpha\beta}$ are irreducible while the tensors $D'^{\alpha\beta}$
and $X'^{\alpha\beta}$ are not. A new pair of tensor can be constructed
in which the trace of $D'^{\alpha\beta}$ and $X'^{\alpha\beta}$
is removed\begin{align}
D^{\alpha\beta}:= & -\icu\left(D'^{\alpha\beta}-\mathcal{L}_{+}g^{\alpha\beta}\right)\\
X^{\alpha\beta}:= & -\icu\left(X'^{\alpha\beta}-\mathcal{L}_{-}g^{\alpha\beta}\right).\end{align}
Obviously they are trace-free by construction,\begin{align}
D_{\blank{\mu}\mu}^{\mu}= & 0\\
X_{\blank{\mu}\mu}^{\mu}= & 0,\end{align}
so $D^{\alpha\beta}$ and $X^{\alpha\beta}$ are irreducible tensors
of total valence two. The tensors $D^{\alpha\beta}$ and $X^{\alpha\beta}$
are also equal to the imaginary parts of $D'^{\alpha\beta}$ and $X'^{\alpha\beta}$
respectively, that is

\begin{align}
D^{\alpha\beta}= & \Im\left\{ D'^{\alpha\beta}\right\} \\
X^{\alpha\beta}= & \Im\left\{ X'^{\alpha\beta}\right\} .\end{align}
and so $D^{\alpha\beta}$ and $X^{\alpha\beta}$ are purely real,

\begin{align}
\Im\left\{ D^{\alpha\beta}\right\} = & 0\\
\Im\left\{ X^{\alpha\beta}\right\} = & 0.\end{align}
 However, in contrast to $T^{\alpha\beta}$ and $Q^{\alpha\beta}$
which are symmetric, $D^{\alpha\beta}$ and $X^{\alpha\beta}$ are
antisymmetric,\begin{align}
D^{\alpha\beta}= & -D^{\beta\alpha}\\
X^{\alpha\beta}= & -X^{\beta\alpha}.\end{align}
Furthermore, $D^{\alpha\beta}$ and $X^{\alpha\beta}$ are mutually
dual, that is\begin{equation}
\dual D^{\alpha\beta}=+X^{\alpha\beta},\quad\dual X^{\alpha\beta}=-D^{\alpha\beta}\label{eq:DXduals}\end{equation}
while $T^{\alpha\beta}$ and $Q^{\alpha\beta}$ are not mutually dual.
In fact, their duals are zero,\begin{align}
\dual T^{\alpha\beta}= & 0\label{eq:Tdual0}\\
\dual Q^{\alpha\beta}= & 0.\label{eq:Qdual0}\end{align}

In a local, Lorentzian metric, coordinate system, the components of
$D^{\alpha\beta}$ and $X^{\alpha\beta}$ are\begin{align}
D^{00}= & 0\\
D^{i0}= & -\Im\left\{ \CompConj{\mathbf{E}}\times\mathbf{B}\right\} \\
D^{ij}= & -\Im\left\{ \mathbf{\CompConj{E}}\otimes\mathbf{E}-\CompConj{\mathbf{B}}\otimes\mathbf{B}\right\} =-\frac{\icu}{2}\left(\mathbf{\CompConj{E}}\times\mathbf{E}-\mathbf{\CompConj{B}}\times\mathbf{B}\right)\times\\
D^{ji}= & -D^{ij}\end{align}
\begin{align}
X^{00}= & 0\\
X^{i0}= & \frac{\icu}{2}\left(\mathbf{\CompConj{E}}\times\mathbf{E}-\mathbf{\CompConj{B}}\times\mathbf{B}\right)\\
X^{ij}= & \Im\left\{ \mathbf{\CompConj{E}}\otimes\mathbf{B}+\CompConj{\mathbf{B}}\otimes\mathbf{E}\right\} =\frac{\icu}{2}\left(\CompConj{\mathbf{E}}\times\mathbf{B}-\mathbf{E}\times\CompConj{\mathbf{B}}\right)\times\\
X^{ji}= & -X^{ij}\end{align}
where we have used the 3-vector components of $F^{\alpha\beta}$.
Being mutually dual means that $D^{\alpha\beta}$ and $X^{\alpha\beta}$
together contain the same information as the two real 3-vectors $\Im\left\{ \CompConj{\mathbf{E}}\times\mathbf{B}\right\} $
and $\icu\left(\mathbf{\CompConj{E}}\times\mathbf{E}-\mathbf{\CompConj{B}}\times\mathbf{B}\right)/2$.
The vector $\Im\left\{ \CompConj{\mathbf{E}}\times\mathbf{B}\right\} $
is recognized to be proportional to the imaginary part of the complex
Poynting vector, see \cite{Jackson99}.

\section{Irreducible concomitants of valence four}

The two tensors, $T'^{\alpha\beta\gamma\delta}$ and $Q'^{\alpha\beta\gamma\delta}$
are completely reducible as follows\begin{align}
T'^{\alpha\beta\gamma\delta}= & 2T^{[\alpha[\delta}g^{\gamma]\beta]}-\frac{\icu}{4}\left(T^{\alpha\mu}\epsilon_{\mu}^{\blank{\mu}\beta\gamma\delta}-T^{\beta\mu}\epsilon_{\mu}^{\blank{\mu}\beta\gamma\delta\alpha}-T^{\gamma\mu}\epsilon_{\mu}^{\blank{\mu}\delta\alpha\beta}+T^{\delta\mu}\epsilon_{\mu}^{\blank{\mu}\alpha\beta\gamma}\right)\\
Q'^{\alpha\beta\gamma\delta}= & 2Q^{[\alpha[\delta}g^{\gamma]\beta]}-\frac{\icu}{4}\left(Q^{\alpha\mu}\epsilon_{\mu}^{\blank{\mu}\beta\gamma\delta}-Q^{\beta\mu}\epsilon_{\mu}^{\blank{\mu}\beta\gamma\delta\alpha}-Q^{\gamma\mu}\epsilon_{\mu}^{\blank{\mu}\delta\alpha\beta}+Q^{\delta\mu}\epsilon_{\mu}^{\blank{\mu}\alpha\beta\gamma}\right),\end{align}
where we are using the usual square bracket notation for indices to
denote antisymmetrization over the enclosed indices (e.g. $T^{\alpha[\delta}g^{\gamma]\beta}=\frac{1}{2}\left(T^{\alpha\delta}g^{\gamma\beta}-T^{\alpha\gamma}g^{\delta\beta}\right)$)
and employing the rule that nested brackets are not operated on by
enclosing brackets (e.g. $T^{[\alpha[\delta}g^{\gamma]\beta]}=\frac{1}{4}\left(T^{\alpha\delta}g^{\gamma\beta}-T^{\alpha\gamma}g^{\delta\beta}-T^{\beta\delta}g^{\gamma\alpha}+T^{\beta\gamma}g^{\delta\alpha}\right)$). 

The other two tensors, $D'^{\alpha\beta\gamma\delta}$ and $X'^{\alpha\beta\gamma\delta}$,
are however not completely reducible. So from them we can construct
two new valence-4 tensors\begin{align}
D^{\alpha\beta\gamma\delta}:= & D'^{\alpha\beta\gamma\delta}-2\icu D^{[\alpha[\delta}g^{\gamma]\beta]}-\frac{2}{3}L_{+}g^{\alpha[\delta}g^{\gamma]\beta}-\frac{1}{3}L_{-}\epsilon^{\alpha\beta\gamma\delta}\\
X^{\alpha\beta\gamma\delta}:= & X'^{\alpha\beta\gamma\delta}-2\icu X^{[\alpha[\delta}g^{\gamma]\beta]}-\frac{2}{3}L_{-}g^{\alpha[\delta}g^{\gamma]\beta}+\frac{1}{3}L_{+}\epsilon^{\alpha\beta\gamma\delta}\end{align}
 both of which fulfill (\ref{eq:4Traceless}), (\ref{eq:4OrthPseu}),
and (\ref{eq:4Irred13}) and are therefore irreducible. Both $\mathcal{D}^{\alpha\beta\gamma\delta}$
and $\mathcal{X}^{\alpha\beta\gamma\delta}$ are real. They are, however,
not independent of each other; in fact\begin{equation}
X^{\alpha\beta\gamma\delta}=\frac{1}{2}\epsilon_{\blank{\alpha}\blank{\beta}\mu\nu}^{\alpha\beta}D^{\mu\nu\gamma\delta},\quad D^{\alpha\beta\gamma\delta}=\frac{1}{2}\epsilon_{\blank{\alpha}\blank{\beta}\mu\nu}^{\alpha\beta}X^{\mu\nu\gamma\delta},\label{eq:XD4dual}\end{equation}
that is, they are mutually dual over their two leftmost indices.

In a local, Lorentzian metric, coordinate system, the components of
$D^{\alpha\beta\gamma\delta}$ and $X^{\alpha\beta\gamma\delta}$
are\begin{align}
D^{\alpha\beta\gamma\delta}\leftrightarrow & D^{AB}=\frac{1}{2}\left(\begin{array}{cc}
\Re\left\{ \CompConj{\mathbf{E}}\otimes\mathbf{E}-\CompConj{\mathbf{B}}\otimes\mathbf{B}\right\} -\frac{2}{3}\mathcal{L}_{+}\mathbf{1}_{3} & \Re\left\{ \CompConj{\mathbf{E}}\otimes\mathbf{B}+\CompConj{\mathbf{B}}\otimes\mathbf{E}\right\} -\frac{2}{3}\mathcal{L}_{-}\mathbf{1}_{3}\\
\Re\left\{ \CompConj{\mathbf{E}}\otimes\mathbf{B}+\CompConj{\mathbf{B}}\otimes\mathbf{E}\right\} -\frac{2}{3}\mathcal{L}_{-}\mathbf{1}_{3} & -\Re\left\{ \CompConj{\mathbf{E}}\otimes\mathbf{E}-\CompConj{\mathbf{B}}\otimes\mathbf{B}\right\} +\frac{2}{3}\mathcal{L}_{+}\mathbf{1}_{3}\end{array}\right)\end{align}
\begin{align}
X^{\alpha\beta\gamma\delta}\leftrightarrow & X^{AB}=\frac{1}{2}\left(\begin{array}{cc}
-\Re\left\{ \CompConj{\mathbf{E}}\otimes\mathbf{B}+\CompConj{\mathbf{B}}\otimes\mathbf{E}\right\} +\frac{2}{3}\mathcal{L}_{-}\mathbf{1}_{3} & \Re\left\{ \CompConj{\mathbf{E}}\otimes\mathbf{E}-\CompConj{\mathbf{B}}\otimes\mathbf{B}\right\} -\frac{2}{3}\mathcal{L}_{+}\mathbf{1}_{3}\\
\Re\left\{ \CompConj{\mathbf{E}}\otimes\mathbf{E}-\CompConj{\mathbf{B}}\otimes\mathbf{B}\right\} -\frac{2}{3}\mathcal{L}_{+}\mathbf{1}_{3} & \Re\left\{ \CompConj{\mathbf{E}}\otimes\mathbf{B}+\CompConj{\mathbf{B}}\otimes\mathbf{E}\right\} -\frac{2}{3}\mathcal{L}_{-}\mathbf{1}_{3}\end{array}\right)\end{align}
when expressed in bivector indexing.

\section{Irreducible tensorial set}

The irreducible tensors constructed in the previous section can be
assembled into an real-irreducible tensorial set in the same spirit
as Fano and Racah \cite{Fano59}. Such a set is given in Table \ref{tab:concomSum02}.
They fulfill all the criteria itemized in Section \ref{sec:Criteria}.

\begin{table}

\caption{\label{tab:concomSum02}Summary of a complete set of irreducible
tensors. The heading {}``No. indep. comp.'' stands for {}``number
of independent components'' . Note that the quantities in parentheses
are alternatives elements of the set.}

\begin{tabular}{ccc}
\hline 
Concomitant&
Total valence&
No. indep. comp.\tabularnewline
\hline
\hline 
$\mathcal{L}_{+}$&
0&
1\tabularnewline
$\mathcal{L}_{-}$&
0&
1\tabularnewline
$T^{\alpha\beta}$&
2&
9\tabularnewline
$\icu Q^{\alpha\beta}$&
2&
9\tabularnewline
$\icu D^{\alpha\beta}$ (or $\icu X^{\alpha\beta}$ )&
2&
6\tabularnewline
$D^{\alpha\beta\gamma\delta}$(or $X^{\alpha\beta\gamma\delta}$ )&
4&
10\tabularnewline
\hline
\end{tabular}
\end{table}

The tensors in Table \ref{tab:concomSum02} form a complete set by
which we mean that any covariant hermitian form tensor concomitant
of $F^{\alpha\beta}$ can be written as linear combinations (including
arbitrary contractions) of $g^{\alpha\beta}$, $g^{\alpha\beta}g^{\gamma\delta}$,
$\epsilon^{\alpha\beta\gamma\delta}$, and the set of tensors in Table
\ref{tab:concomSum02}. In terms of the number of independent components,
the tensors in the table constitute 36 independent components in total.
This is exactly the number of unique bilinear combinations possible
from the 6 complex components of the bivector if both real and imaginary
parts are counted separately. In the real-valued bivector case, there
are only 21 second-order combinations possible in total. This matches
the total number of independent components in the set: $\mathcal{L}_{+}$,
$\mathcal{L}_{-}$, $T^{\alpha\beta}$, $D^{\alpha\beta\gamma\delta}$
(or $X^{\alpha\beta\gamma\delta}$).

Note that a complete irreducible set of tensors bilinear in $F^{\alpha\beta}$
is not entirely unique. This is because, tensor $D^{\alpha\beta}$
is isomorphic with $X^{\alpha\beta}$ (its dual) and $D^{\alpha\beta\gamma\delta}$
is isomorphic with $X^{\alpha\beta\gamma\delta}$, that is, either
$X^{\alpha\beta}$ or $X^{\alpha\beta\gamma\delta}$ could be used
in place of $D^{\alpha\beta}$ or $D^{\alpha\beta\gamma\delta}$ respectively.
Therefore the complete irreducible tensorial set in Table \ref{tab:concomSum02}
is but one out of four of such sets possible.

We have throughout assumed that $F^{\alpha\beta}$ is an arbitrary
complex bivector. If $F^{\alpha\beta}$ were real, then $\icu Q^{\alpha\beta}$,
$\icu D^{\alpha\beta}$, $\icu X^{\alpha\beta}$ are all zero; in
other words, all the tensors in the Table with coefficient $\icu$
vanish in the ordinary case of real-valued bivectors, leaving only
$\mathcal{L}_{+}$, $\mathcal{L}_{-}$, $T^{\alpha\beta}$, $D^{\alpha\beta\gamma\delta}$,
$X^{\alpha\beta\gamma\delta}$. Thus the existence of $\icu Q^{\alpha\beta}$,
$\icu D^{\alpha\beta}$, $\icu X^{\alpha\beta}$ is a direct consequence
of the fact that an arbitrary complex $F^{\alpha\beta}$ can have
an imaginary part.

\section{Conclusion}

We have derived a complete set of irreducible tensorial concomitants
that are bilinear in an arbitrary complex bivector. The set is summarized
in Table \ref{tab:concomSum02}. The tensors in the set are all hermitian
forms in the complex bivector $F^{\alpha\beta}$ and invariant, up
to a sign, under the duality transformation of the bivector. That
the valence-2 tensors are irreducible is equivalent them being trace-free,
see Eq. (\ref{eq:2Traceless}), and that the valence-4 tensors are
irreducible is equivalent to saying that they simultaneously fulfill
the conditions (\ref{eq:4Traceless}), (\ref{eq:4OrthPseu}), and
(\ref{eq:4Irred13}).

Of the tensors in the Table, $\mathcal{L}_{+}$,$\mathcal{L}_{-}$,$T^{\alpha\beta}$
are all hermitian form generalizations of previously known bilinear
concomitants. All other tensors are, as far as the authors are aware,
novel.

\begin{acknowledgments}
This work was sponsored by PPARC ref: PPA/G/S/1999/00466 and PPA/G/S/2000/00058.
\end{acknowledgments}

\end{document}